\documentclass[aps,prl,twocolumn,superscriptaddress,showpacs,altaffillsymbol,nofootinbib]{revtex4-1}
\usepackage{amssymb}
\usepackage{amsmath}
\usepackage{commath}
\usepackage{graphicx}
\usepackage{verbatim}
\usepackage{epsfig}
\usepackage{float}
\usepackage{epstopdf}
\usepackage{color}
\usepackage{siunitx}
\usepackage{array}
\usepackage{physics}
\newcolumntype{P}[1]{>{\centering\arraybackslash}p{#1}}
\newcolumntype{M}[1]{>{\centering\arraybackslash}m{#1}}

\begin{document}


\title{Low Energy Magneto-optics of Tb$_{2}$Ti$_{2}$O$_{7}$ in [111] Magnetic Field}

\author{Xinshu Zhang}
\affiliation{Institute for Quantum Matter, Department of Physics and Astronomy, The Johns Hopkins University, Baltimore, Maryland 21218, USA}

\author{Yi Luo}
\affiliation{Institute for Quantum Matter, Department of Physics and Astronomy, The Johns Hopkins University, Baltimore, Maryland 21218, USA}

\author{T. Halloran}
\affiliation{Institute for Quantum Matter, Department of Physics and Astronomy, The Johns Hopkins University, Baltimore, Maryland 21218, USA}

\author{J. Gaudet}
\affiliation{Institute for Quantum Matter, Department of Physics and Astronomy, The Johns Hopkins University, Baltimore, Maryland 21218, USA}

\author{Huiyuan Man}
\affiliation{Institute for Quantum Matter, Department of Physics and Astronomy, The Johns Hopkins University, Baltimore, Maryland 21218, USA}

\author{S. M. Koohpayeh}
\affiliation{Institute for Quantum Matter, Department of Physics and Astronomy, The Johns Hopkins University, Baltimore, Maryland 21218, USA}
\affiliation{Department of Materials Science and Engineering, The Johns Hopkins University, Baltimore, Maryland 21218, USA}

\author{N. P. Armitage}
\affiliation{Institute for Quantum Matter, Department of Physics and Astronomy, The Johns Hopkins University, Baltimore, Maryland 21218, USA}

\date{\today}

\begin{abstract}
The pyrochlore magnet Tb$_{2}$Ti$_{2}$O$_{7}$ shows a lack of magnetic order to low temperatures and is considered to be a quantum spin liquid candidate.   We perform time-domain THz spectroscopy on high quality Tb$_{2}$Ti$_{2}$O$_{7}$ crystal and study the low energy excitations as a function of [111] magnetic field with high energy resolution. The low energy crystal field excitations change their energies anomalously under magnetic field. Despite several sharp field dependent changes, we show that the material's spectrum can be described not by a phase transitions, but by field dependent hybridization between the low energy crystal field levels.  We highlight the strong coupling between spin and lattice degrees of freedom in Tb$_{2}$Ti$_{2}$O$_{7}$ as evidenced by the magnetic field tunable crystal field environment. Calculations based on single ion physics with field induced symmetry reduction of the crystal field environment can reproduce our data.
\end{abstract}
\maketitle

Geometric frustration is a central theme in quantum magnets as it suppresses long range magnetic order and frequently causes interesting low temperature states~\cite{balents2010balents}. Rare earth pyrochlores with corner sharing tetrahedra of magnetic ions have strong geometric frustrations.  They have been subjects of intense interest for many years~\cite{RevModPhys.82.53,ramirez1999zero}. Famous examples include canonical spin ice compound Dy$_{2}$Ti$_{2}$O$_{7}$ and Ho$_{2}$Ti$_{2}$O$_{7}$, where at low temperature spins follow a two-in-two-out ``ice rule" on the tetrahedra and the low energy magnetic excitations are magnetic monopoles connected by a Dirac string~\cite{fennell2009magnetic,bramwell2001spin,castelnovo2008magnetic,morris2009dirac}. Quantum spin liquids (QSLs) are states where spins do not order down to zero temperature and remain dynamic~\cite{RevModPhys.89.025003,savary2016quantum}. They have attracted attention for decades due to the potential for exotic quasiparticles as well as potential applications to quantum computation~\cite{banerjee2016proximate,kitaev2003fault,paddison2017continuous,han2012fractionalized}.  Yb$_{2}$Ti$_{2}$O$_{7}$ was once deemed to be quantum spin ice (QSI), but recently  experiments on high quality single crystals suggest a coexistence and competition of ferro and anti-ferromagnetism~\cite{scheie2019multiphase,ross2011quantum,PhysRevLett.119.127201,pan2016measure,PhysRevLett.115.267208,PhysRevB.92.064425}. It is believed that Tb$_{2}$Ti$_{2}$O$_{7}$ may be a QSL~\cite{PhysRevLett.82.1012}. Unlike Yb$_{2}$Ti$_{2}$O$_{7}$ that orders at 270 mK, Tb$_{2}$Ti$_{2}$O$_{7}$ shows no magnetic order down to 50 mK despite an anti-ferromagnetic Curie-Weiss temperature $\theta_{CW}$ $\approx$ 19 K~\cite{arpino2017impact,PhysRevB.62.6496}. Neutron scattering shows power law correlations~\cite{PhysRevLett.109.017201}, that bear resemblance to the pinch points found in classical spin ice. $\mu$SR shows persistent spin fluctuations down to 70 mK with a relaxation rate 0.04 THz, supporting spin liquid behavior~\cite{PhysRevLett.82.1012}. 

Tb$_{2}$Ti$_{2}$O$_{7}$ is unique among the pyrochlore titanate family for having particularly low energy crystal electric field levels (CEF) and a strong spin-lattice coupling, which are believed to be important for the spin liquid physics. The trigonal symmetry CEF acting on the Tb$^{3+}$ ion leads to a low energy first excited doublet that is separated from the ground state doublet by $\Delta \sim$ 0.4 THz (1.6 meV)~\cite{PhysRevB.94.024430,PhysRevB.89.134410,PhysRevBcef,PhysRevB.77.214310,PhysRevB.62.6496,PhysRevB.94.024430,PhysRevB.89.134410,PhysRevBcef}. This is in contrast to other pyrochlore titanates, where the lowest excited CEF level is well isolated from the ground state~\cite{PhysRevB.62.6496}. Virtual crystal field excitation scenarios and upper branch magnetism have been proposed for Tb$_{2}$Ti$_{2}$O$_{7}$ to account for spin liquid physics~\cite{PhysRevLett.98.157204,PhysRevB.99.224407}.  Moreover, because Tb$^{3+}$ is a non-Kramers ion, Tb$_{2}$Ti$_{2}$O$_{7}$ is susceptible to distortions. Although the presence of static distortions have been debated~\cite{PhysRevB.84.140402,PhysRevB.84.184409,PhysRevB.78.094418,rule2009tetragonal}, many experiments have demonstrated the strong coupling between spin and lattice degrees of freedom.  For instance, giant magnetostriction~\cite{klekovkina2014crystal}, a coupling between CEF levels and a transverse phonon~\cite{fennell2014magnetoelastic,constable2017double,PhysRevB.99.224431}, dynamic structural fluctuations as well as a structural transition in pulsed magnetic field have been reported in Tb$_{2}$Ti$_{2}$O$_{7}$~\cite{PhysRevLett.99.237202,PhysRevLett.105.077203}, suggesting the important role of the lattice.  
Recently, Raman scattering measurements have reported anomalous CEF excitations with [111] magnetic field, which have been interpreted as a structural phase transition $\sim$2.5 T, leading to electric dipoles induced by magnetic monopoles~\cite{PhysRevLett.124.087601}. 

\begin{figure*}
	\centering
	\includegraphics[width=1.98\columnwidth]{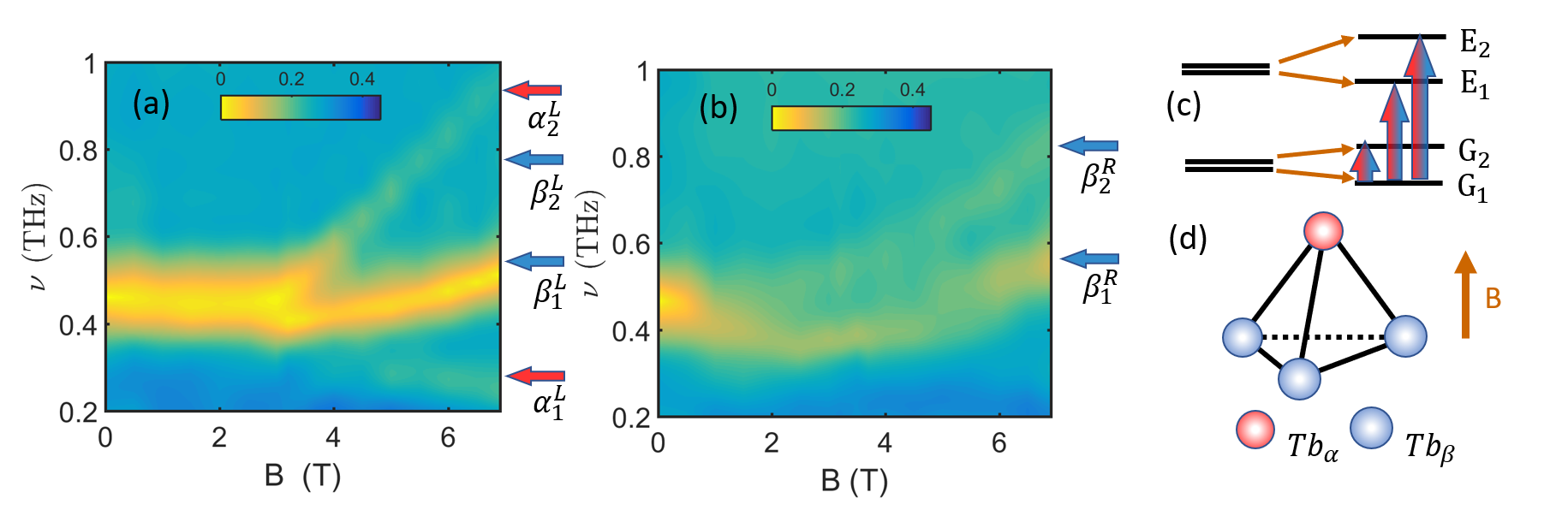}
	\caption{ Transmission amplitude for \textbf{(a)} left and \textbf{(b)} right circularly polarized THz light as a function of frequency  and magnetic field up to 6.9 T at 1.6 K. Four excitations are observed in left channel indicated by $\alpha_{1}^{L}$, $\beta_{1}^{L}$,$\beta_{2}^{L}$, $\alpha_{2}^{L}$ and two excitations in right channel indicated by $\beta_{1}^{R}$ and $\beta_{2}^{R}$.  \textbf{(c)}  schematic of CEF splitting in magnetic field labelled as G$_{1}$, G$_{2}$, E$_{1}$ and E$_{2}$. The red and blue arrows indicate transitions at two different sites $\alpha$ and $\beta$.  \textbf{(d)} Tetrahedron with red and blue spheres represent the two different Tb ions under [111] magnetic field.  }
	\label{fig:1}
\end{figure*} 

In this work, we perform time-domain THz spectroscopy (TDTS) on high quality single crystal Tb$_{2}$Ti$_{2}$O$_{7}$ to study the evolution of the low energy crystal field excitations as function of magnetic field along [111] direction.  We observed several crystal field anomalies.  Although the sharpness of these anomalies might have been taken to be evidence for a field dependent phase transition, we show that such an interpretation is not necessary.  Field dependent changes to the CEF environment can capture the essential features of the experiment.

TDTS measurements were performed on optically clear traveling solvent floating zone grown crystals in a custom-built polarization modulation set-up with a frequency range from 0.2 to 2 THz (0.83-8.3 meV)~\cite{morris2014hierarchy} that all components of the transmission tensor (T$_{xx}$, T$_{xy}$) can be measured. The complex THz transmission of a 3$\times$3$\times$0.15 mm$^3$  single crystal was measured down to 1.6 K  with external fields up to 6.9 T in the Faraday ($\bf{k}\!\parallel\!\bf{H_{dc}}$)  geometry, where $\bf{k}$ is the direction of light propagation.  Under magnetic field, the linearly polarized THz pulse rotates and becomes elliptically polarized as it passes through the sample due to the Faraday rotation. The complex transmission in the Faraday geometry in the linear basis can be converted to the circular basis via the expression $T_{R,L} = T_{xx} \pm i T_{xy}$~\cite{PhysRevX.8.031001,pan2014low}.  The circular basis are eigenpolarization for transmission in a cubic lattice under magnetic field.  The THz transmission in a thick single crystal is a measure of the complex susceptibility~\cite{PhysRevX.8.031001,laurita2015singlet}.  Fig. 1(a) and (b) show transmission magnitudes as a function of frequency and magnetic field at a temperature of 1.6 K for left and right circularly polarized light THz light. A number of excitations are seen as bright yellow features, which correspond to CEF excitations. The fact that different excitations are observed in the two channels correspond to their distinct selection rules.

\begin{figure*}
	\centering
	\includegraphics[width=1.98\columnwidth]{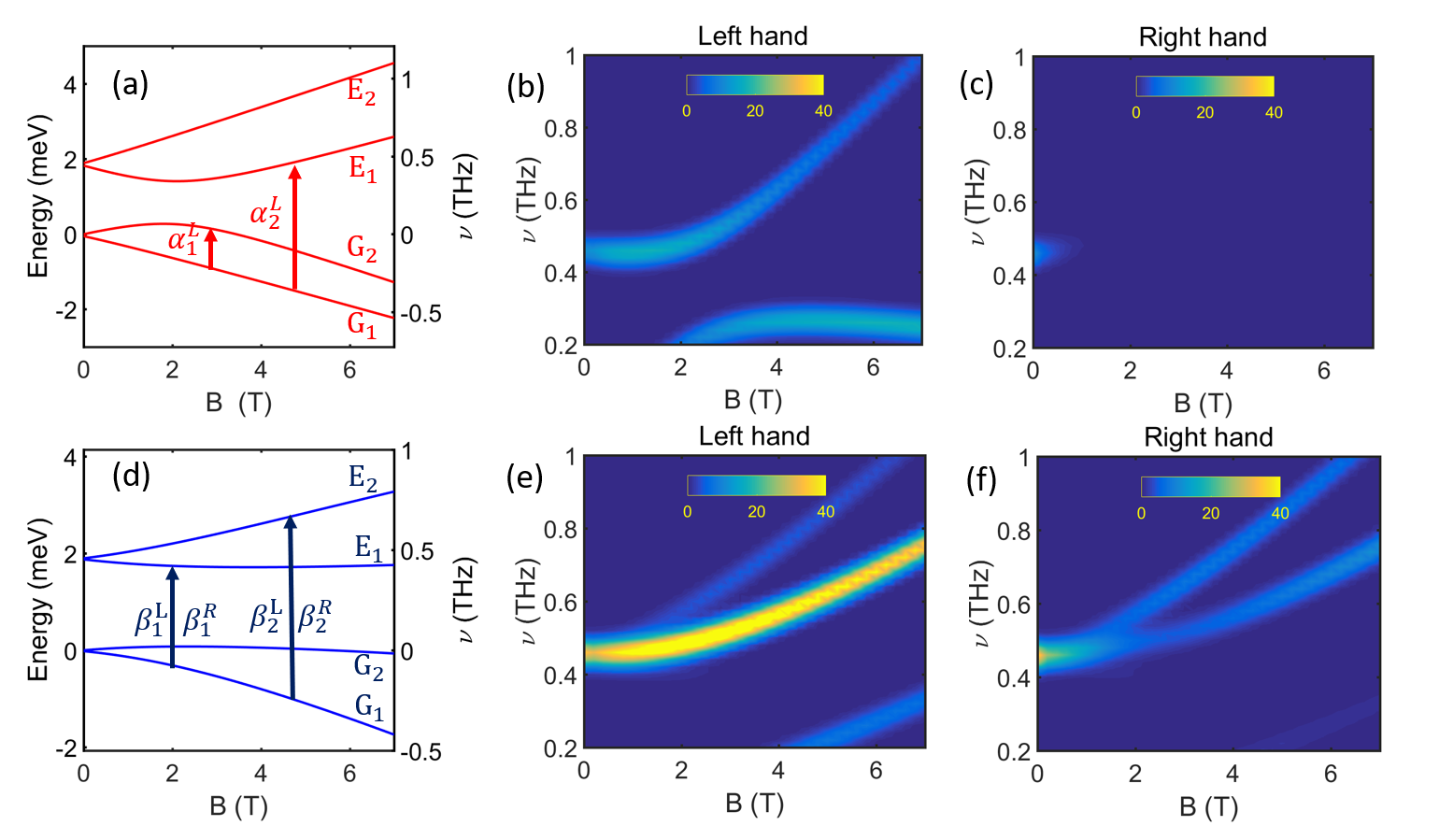}
	\caption{  \textbf{(a)} Calculated energy of four lowest crystal field as a function of field for Tb$_{\alpha}$. The observed excitations $\alpha_{1}^{L}$ and $\alpha_{2}^{L}$ correspond to transition  $G_{1}$  $\rightarrow$ $G_{2}$ and  $G_{1}$  $\rightarrow$ $E_{1}$ of Tb$_{\alpha}$.   Colour plot of calculated intensity as function of energy and field for Tb$_{\alpha}$ in left \textbf{(b)} and right \textbf{(c)} channel.  The weak intensity at low field in \textbf{(c)} is transition from thermally populated $G_{2}$  $\rightarrow$ $E_{2}$.   \textbf{(d)} Calculated energy of four lowest crystal field as a function of field for Tb$_{\beta}$. The observed excitations $\beta_{1}^{L}$,$\beta_{1}^{R}$ and $\beta_{2}^{L}$,$\beta_{2}^{R}$ correspond to transition  $G_{1}$  $\rightarrow$ $E_{1}$ and  $G_{1}$  $\rightarrow$ $E_{2}$ of Tb$_{\beta}$, respectively. \textbf{(e-f)} Colour plot of calculated intensity as function of energy and field for Tb$_{\beta}$ in left and right.}
	\label{fig:2}
\end{figure*} 

The observed excitations are expected to originate from transitions from the ground state doublet ($G_{1}$ and $G_{2}$) to the first excited state doublet ($E_{1}$ and $E_{2}$), as higher excited states are at much larger energies~\cite{PhysRevB.94.024430,PhysRevB.89.134410,PhysRevBcef,PhysRevB.77.214310}.  In zero magnetic field, in the basis of $\vert m \rangle$=$\vert J$=$6 ;J_{z}$=$m \rangle$ the ground state doublet is approximately $\psi_{G_{1,2}} \approx a_{G}\vert \pm 4 \rangle \pm b_{G}\vert \mp 5 \rangle$ and the first excited state is approximately $\psi_{E_{1,2}} \approx a_{E}\vert \pm 5 \rangle \pm b_{E} \vert \mp 4 \rangle$. Applying magnetic field along [111], the ground and first excited state doublets split. Because the experiments were performed at 1.6 K ($\approx $ 0.03 THz and 0.13 meV) the thermal population of all higher levels in high field are negligible and hence only the transitions starting from the ground state G$_{1}$ are important.  There are two distinct Tb sites ($\alpha$ and $\beta$) differentiated by their alignment to the [111] field as shown in Fig. 1(d). Hence, there are 6 different transitions expected, but in experiment not all of them can be seen due to optical selection rules, experimental resolution and limited THz range, etc.  We observe four features in the left channel Fig. 1(a) ($\alpha_{1}^{L}$, $\beta_{1}^{L}$, $\beta_{2}^{L}$, $\alpha_{2}^{L}$) and two in the right channel Fig. 1(b) ($\beta_{1}^{R}$, $\beta_{2}^{R}$).  Some of the features appear more clearly in the ellipticity as shown in the SI.  As discussed below, features $\beta_{1}^{L}$/ $\beta_{1}^{R}$ and $\beta_{2}^{L}$/$\beta_{2}^{R}$ correspond to the same transitions, which appears in both channels.  $\alpha_{1}^{L}$ and $\alpha_{2}^{L}$ only appear in the L channel due to selection rules.  Other notable aspects of the data include an almost constant CEF excitation $\alpha_{2}^{L}$ up to 3 T, which then changes its behavior abruptly near 3 T, a behavior that is in stark contrast to the common Zeeman splitting, and a seemingly ``new" excitation $\alpha_{1}^{L}$  that shows increasing intensity with magnetic field.

To gain further insight on this data, we performed calculations based on single ion physics for D$_{3d}$ point symmetry for Tb$^{3+}$.  The effective Hamiltonian in terms of Stevens operators $O^{q}_{k}$, which applies for a basis comprised of a single $J$ level ($J$=6 for Tb$^{3+}$) in the LS coupling scheme, is 
\begin{eqnarray}
H_{CEF}=\theta_{2}D_{2}^{0}O_{2}^{0}+\theta_{4}D_{4}^{0}O_{4}^{0}+\theta_{4}D_{4}^{3}O_{4}^{3} \nonumber\\+\theta_{6}D_{6}^{0}O_{6}^{0}+\theta_{6}D_{6}^{3}O_{6}^{3}+\theta_{6}D_{6}^{6}O_{6}^{6},
\end{eqnarray} where $\theta_{k}$ are the reduced matrix elements.  Crystal field parameters $B^{k}_{q}$ are obtained by fitting to zero field neutron crystal field energies with the $k$ averaged lowest excitation at 1.6 meV replaced by $\sim$1.9 meV at the $\Gamma$ point as observed in THz spectroscopy~\cite{PhysRevB.94.024430}.  $D^{q}_{k}$ and $B^{k}_{q}$ are related by $D^{q}_{k}=\lambda^{q}_{k}B^{k}_{q}$~\cite{Hutchings_1964}. The obtained $B^{k}_{q}$ parameters are given in SI and are close to previous works~\cite{PhysRevB.94.024430}.  $\theta_{k}$ and $\lambda^{q}_{k}$ values are also given in the SI.

In order to account for the evolution of crystal field excitations as function of magnetic field, we include a Zeeman term \begin{equation}
H_{Zeeman}=-\mu_{B}g_{J}\vec{B}\cdot\vec{J}.
\end{equation} 
By diagonalizing the Hamiltonian, we obtain the eigenenergies and eigenfunctions at each magnetic field.  We calculate the spectra for Tb$_{\alpha}$ and Tb$_{\beta}$ separately because the Zeeman term is different due to relative orientation of field with respect to the local [111] axis.

\begin{figure*}
	\centering
	\includegraphics[width=1.98\columnwidth]{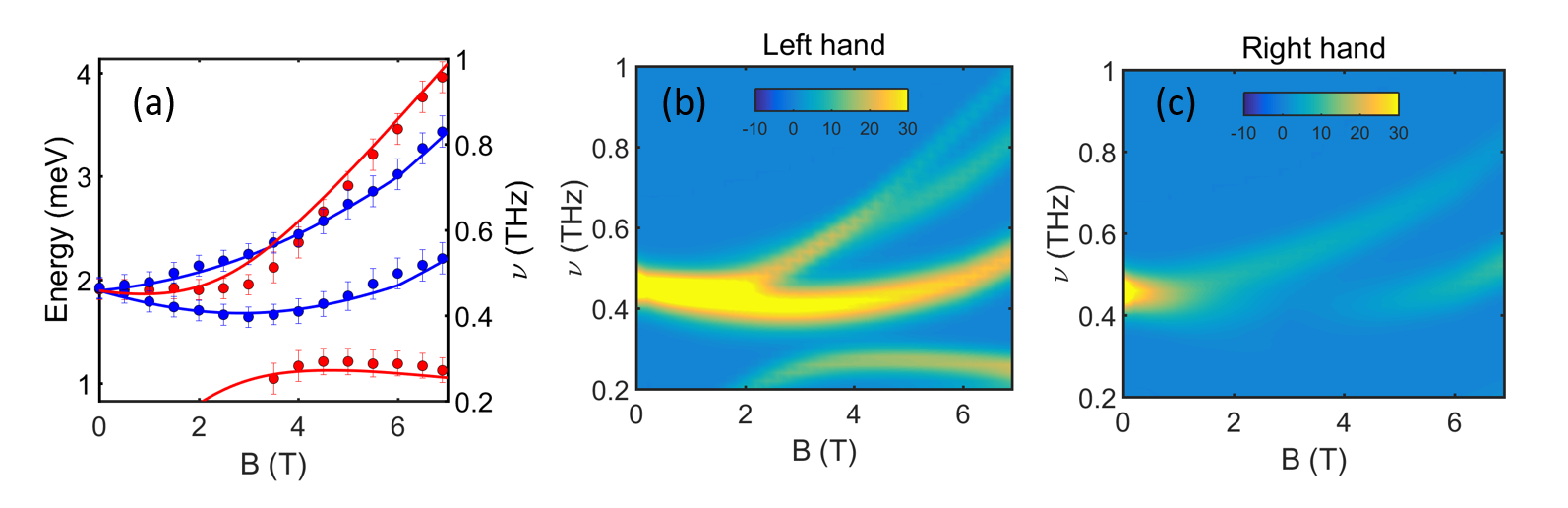}
	\caption{  \textbf{(a)} Comparison between experiment and calculated energy with consideration of field induced crystal field environment change. Red and blue represent Tb$_{\alpha}$ and Tb$_{\beta}$.  Colour plot of calculated intensity as function of energy and field for Tb$_{\alpha}$ and Tb$_{\beta}$ in left \textbf{(b)} and right \textbf{(c)} channel. }
	\label{fig:3}
\end{figure*}

We plot the field evolution of the four lowest energy levels for Tb$_{\alpha}$ and Tb$_{\beta}$ as shown in Fig. 2(a) and Fig. 2(d), respectively. The transitions between different states correspond closely to experimentally observed excitations. The CEF excitations couple to the magnetic field of the THz pulse.  The general selection rule for magnetic dipole excitations is  $\Delta m=0,\pm1$ and $\Delta J=0,\pm1$ with no parity change, which is already satisfied by the $E_{g}$ symmetry of ground and first excited doublets.  In the Faraday geometry with $J$=6 for Tb$^{3+}$, because of the conservation of angular momentum, excitations with $\Delta m=+1,-1$  should be seen in left and right channels, respectively. $\Delta m=0$ is forbidden in the Faraday geometry.  Based on selection rules, the transition  $G_{1}$  $\rightarrow$ $G_{2}$ and $G_{1}$  $\rightarrow$ $E_{1}$ for Tb$_{\alpha}$ have $\Delta m=+1$ and should only be seen in the left channel as $\alpha^{L}_{1}$ and $\alpha^{L}_{2}$. $G_{1}$  $\rightarrow$ $E_{2}$ which would only be allowed with $\Delta m=0$ cannot be seen in either left or right channels for Tb$_{\alpha}$.  Fig. 2(b) and (c) are color plots of computed spectra as function of magnetic field for Tb$_{\alpha}$. The intensity is calculated based on Fermi's golden rule (SI).  For Tb$_{\beta}$ where the field is not along the local [111] direction and transverse components of the field makes the wavefunction a mix of $\vert \pm 4 \rangle$ and $\vert \pm 5 \rangle$, transitions at the $\beta$ sites composed of $\Delta m=\pm 1$  can be seen in both left and right channel as shown in Fig.  2(e) and (f). For Tb$_{\beta}$,   $G_{1}$  $\rightarrow$ $E_{1}$ are observed as $\beta^{L}_{1}$ and $\beta^{R}_{1}$ in left and right channel with similar energies but different intensities.   $G_{1}$  $\rightarrow$ $E_{2}$ are observed as $\beta^{L}_{2}$ and $\beta^{R}_{2}$ as shown in Fig. 1(a) and (b).  

 The number of observed transitions, their intensities, their rough field dependence and even aspects of the anomalies are reproduced in the simple single ion model.  One distinct feature is the kink-like behavior of $\alpha^{L}_{2}$ around 3 T, which might be taken to be indicative of a phase transition.  In fact, Ref.~\cite{PhysRevLett.124.087601} proposed a structural phase transition in a 2.3 T [111] field due to displacement of O$''$ to explain nonlinear phonon splitting under magnetic fields.  Here we have shown this behavior can be understood from energy hybridization within the four lowest crystal energy levels, e.g., Fig. 2(a) shows level repulsion between $G_{2}$ and $E_{1}$ for Tb$_{\alpha}$ , which cause a sharp up-turn of $\alpha^{L}_{2}$ at 3 T as shown in Fig. 2(b). With increasing magnetic field, the intensity of $\alpha^{L}_{1}$ increases  due to an enhanced transition amplitude based on Fermi's golden rule, which looks like a ``new" excitation just appearing at 3 T. In addition, we performed specific heat measurements at 1.6 K (See SI) and found only a broad hump around 0.7 T, which is expected for low energy CEF, but no sharp peak at 2-3 T as expected for structural phase transition. Therefore, we believe those sharp features come from low energy CEF hybridization instead of structural phase transition.

In the above calculation, with only single ion CEF and Zeeman terms involved, one can already explain the experiment reasonably well, but there remains some quantitative inconsistencies, particularly for the Tb$_{\beta}$ contribution. For Tb$_{\beta}$, the transition $G_{1}$  $\rightarrow$ $G_{2}$ should be visible in THz range but we do not observe it. In addition, both of the calculated transitions $G_{1}$  $\rightarrow$ $E_{1}$ and $G_{1}$  $\rightarrow$ $E_{2}$ are predicted to be much higher than the experimental data (see SI).  Clearly, there are higher order effects we have not considered.  We believe that magnetoelastic coupling or exchange interactions are not large effects.   Although we find indirect evidence for magnetoelastic coupling,  a small energy shift for the transition from the ground doublet $G$ to first excited doublet $E$, the magnetoelastic coupling constant is limited by such small energy scale and has little contribution to the calculation (see SI).  Of course an exchange interaction is necessary to reproduce the dispersion of the CEFs, but it does not improve agreement as function of field at the $\Gamma$ point (SI).  Both moderate magnetoelastic couplings and exchange interactions can lead to finite splitting of the zero field doublets, which is inconsistent with an almost invisible splitting in the experiment. This gives an upper bound on their significance.

The preceding analysis treats the CEF Hamiltonian as independent of external magnetic field. However, the magnetic field can influence the crystal field environment as follows.  The orbital angular momentum results from the spatially anisotropic 4$f$ wave functions that can be envisioned as oblate electron charge cloud.  Applying magnetic field along [111] direction aligns the spin of Tb and also the electron cloud because of spin-orbital coupling. This modifies the interaction between valence electrons of Tb and its neighbour ligands (O$''$) via Coulomb repulsion, which results in field induced crystal environment changes. This should be a ubiquitous effect although the influence is generally negligible. Here, such effect may be amplified due to the dynamic vibration of atoms, which makes the overlap between electrons clouds possibly larger.  Therefore we incorporate magnetic field induced crystal field changes in our calculations. Applying a field reduces the symmetry for Tb$_{\beta}$ from D$_{3d}$ to C$_{1}$, which allows additional Stevens operators to appear in $H_{CEF}$. On the other hand, Tb$_{\alpha}$ remains D$_{3d}$, so we keep the same Stevens operators but adding extra field dependence to the original $B_{q}^{k}$. For simplicity, we approximate the dependence of the changes of CEF parameters to be linear on applied field.  We fit the energy excitations to the refined model.  As shown in Fig. 3(a), the experimental spectra can be reproduced essentially perfectly: the calculated energies of $\beta_{1}$ and $\beta_{2}$ for Tb$_{\beta}$ are lower and match with data now; the transition  $G_{1}$  $\rightarrow$ $G_{2}$ is suppressed to below the THz range. The obtained $\Delta B_{q}^{k}$ is relatively small compared to the original $B_{q}^{k}$ and has negligible impact on the high energy excitations. The good agreement between transmission amplitude Fig. 1 and the simulated intensity plots can be seen in Fig. 3 (b-c)  for both left and right channels.

There are a few things to note. (1) For better agreement with experiment, we use a slightly reduced $g$ factor $\sim$1.3 for the calculation instead of the Land'{e} $g_{J}=1.5$. This small correction only modifies the results slightly and does not change the essential features (SI). There are at least two possibilities for the reduced $g$ factor. First, the Land'{e}  result assumes the LS coupling scheme ($J$=6) and ignores higher-order couplings that can reduce $g$.  Moreover, the possibility of a dynamic Jahn-Teller effect has been reported~\cite{PhysRevLett.99.237202,PhysRevB.89.085115} that can lead to a reduced $g$ factor due to quenching of the orbital moment ~\cite{PhysRev.138.A1727}. (2) The coupling between magnetic field and the CEF changes the electron charge cloud and reduces the symmetry of Tb$_{\beta}$ to C$_{1}$. This may lead to a displacement of atoms,  e.g., O$''$, but not necessarily. With the likely dynamic Jahn-Teller effect in Tb$_{2}$Ti$_{2}$O$_{7}$,  atoms may vibrate around the original position without having  finite displacement~\cite{PhysRevLett.99.237202,PhysRevB.89.085115}. (3) The lower energies of Tb$_{\beta}$ (with a reduced Zeeman splitting) can be phenomenologically understood as an induced internal magnetic field compensating the external field.  With applied magnetic field, there is more overlap between the Tb and O electron cloud, which impedes the further rotation of the Tb electron clouds as well as spins  due to spin-orbit coupling. This is equivalent to having an opposite internal field that compensates the external field, hence, suppressing the rotation of spins toward the [111] direction. With increasing the external field further, the spin of Tb$_{\beta}$ gradually aligns with the magnetic field, and so during the fitting we have a threshold of magnetic field $\sim$ 6 T, above which the  $\Delta B_{q}^{k} $  does not change further.

In conclusion, we performed THz spectroscopy on high quality single crystal Tb$_{2}$Ti$_{2}$O$_{7}$ with a [111] magnetic field to study its low energy crystal field excitations. Anomalous crystal field excitations are observed and explained by low energy CEF hybridization. Single ion calculation with inclusion of the field induced changes to the crystal field environment is able to reproduce the spectra  quantitatively. Our work suggests the interplay among multiple degrees of freedom in Tb$_{2}$Ti$_{2}$O$_{7}$ and sets a benchmark for calculating the crystal field excitations of spin liquid in magnetic field. In future experiments, ultrasonic measurements that are sensitive to elastic constant and diffraction measurements that can detect the displacement of O$''$ are desired. The exotic and diverse physics in Tb$_{2}$Ti$_{2}$O$_{7}$ deserves further experimental and theoretical investigations.

The authors would like to thank C. Broholm and T. Fennell for helpful discussions. This work at JHU was supported as part of the Institute for Quantum Matter, an EFRC funded by the DOE-BES under DE-SC0019331.  We would like to thank T. Fennell for helpful correspondence.

\bibliography{references}

\end{document}


\title{Low Energy Magneto-optics of Tb$_{2}$Ti$_{2}$O$_{7}$ in [111] Magnetic Field}

\author{Xinshu Zhang}
\affiliation{Institute for Quantum Matter, Department of Physics and Astronomy, The Johns Hopkins University, Baltimore, Maryland 21218, USA}

\author{Yi Luo}
\affiliation{Institute for Quantum Matter, Department of Physics and Astronomy, The Johns Hopkins University, Baltimore, Maryland 21218, USA}

\author{T. Halloran}
\affiliation{Institute for Quantum Matter, Department of Physics and Astronomy, The Johns Hopkins University, Baltimore, Maryland 21218, USA}

\author{Jonathan Gaudet}
\affiliation{Institute for Quantum Matter, Department of Physics and Astronomy, The Johns Hopkins University, Baltimore, Maryland 21218, USA}

\author{Huiyuan Man}
\affiliation{Institute for Quantum Matter, Department of Physics and Astronomy, The Johns Hopkins University, Baltimore, Maryland 21218, USA}

\author{S. M. Koohpayeh}
\affiliation{Institute for Quantum Matter, Department of Physics and Astronomy, The Johns Hopkins University, Baltimore, Maryland 21218, USA}
\affiliation{Department of Materials Science and Engineering, The Johns Hopkins University, Baltimore, Maryland 21218, USA}

\author{N. P. Armitage}
\affiliation{Institute for Quantum Matter, Department of Physics and Astronomy, The Johns Hopkins University, Baltimore, Maryland 21218, USA}

\date{\today}

\maketitle

\subsection*{1. Sample Synthesis}
The single crystals used in this work were grown with the
traveling-solvent floating zone method ~\cite{arpino2017impact}.  Depending on synthesis parameters and the preparation method used, different quality samples have been prepared with respect to the type/amount of structural disorder (e.g. off-stoichiometry, Tb$^{4+}$ and Ti$^{3+}$ inclusions, stuffing, etc.).  Therefore, growth and development of stoichiometric, pure, and high-structural quality single crystals of Tb$_{2}$Ti$_{2}$O$_{7}$ are crucial, as it was in the case for Yb$_{2}$Ti$_{2}$O$_{7}$~\cite{arpino2017impact}.  Unlike Tb$_{2}$Ti$_{2}$O$_{7}$ float-zone (FZ) grown crystals, the traveling solvent floating zone (TSFZ) technique helps prevent high temperature phase instabilities, and one obtain a clear color and transparent as-grown crystal, confirming the correct oxidation states for both Tb$^{3+}$ and Ti$^{4+}$ cations.  This is in contrast to brownish color of Tb$_{2}$Ti$_{2}$O$_{7}$ grown by traditional floating zone method.  Powder X-ray diffraction confirmed the purity to be identical to the starting stoichiometrically synthesized powder samples~\cite{Koohpayeh20}.

The single crystals have to be polished down to 0.15 mm thickness to show adequate transmission.  A small piece of Tb$_{2}$Ti$_{2}$O$_{7}$ single crystal was used to measure the heat capacity in a Quantum Design PPMS.

\begin{figure}[h]
	\centering
	\includegraphics[width=0.98\columnwidth]{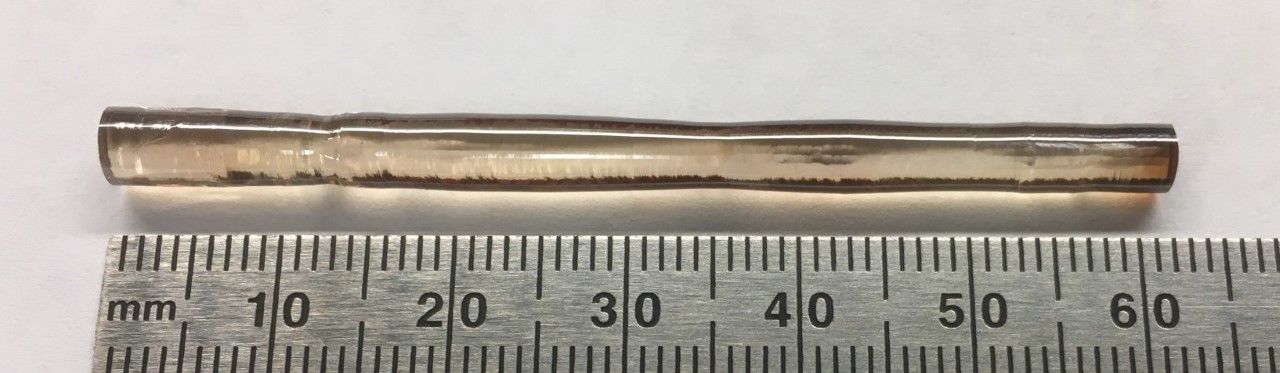}
	\caption{High quality stoichiometric single crystal of  Tb$_{2}$Ti$_{2}$O$_{7}$ grown by the traveling solvent floating zone (TSFZ) technique.}
	\label{fig:S1}
\end{figure}

\subsection*{2. Time-domain THz Spectroscopy (TDTS) Setup and Analysis}
Time-domain THz spectroscopy was performed in a home built setup at Johns Hopkins University. THz pulses with a bandwidth between 0.2 to 2 THz, were generated by a photoconductive antenna (emitter) upon illumination by an infrared laser and then detected by another photoconductive antenna (receiver). The sample was mounted on a 3 mm diameter circular aperture. The electric field profiles of the THz pulses transmitted through the sample and an identical bare aperture were recorded as a function of time and then converted to the frequency domain by Fast Fourier Transforms (FFTs). By dividing the FFTs of the sample and aperture scans, we can obtain the complex transmission of the sample.

\begin{figure}[h]
	\centering
	\includegraphics[width=0.98\columnwidth]{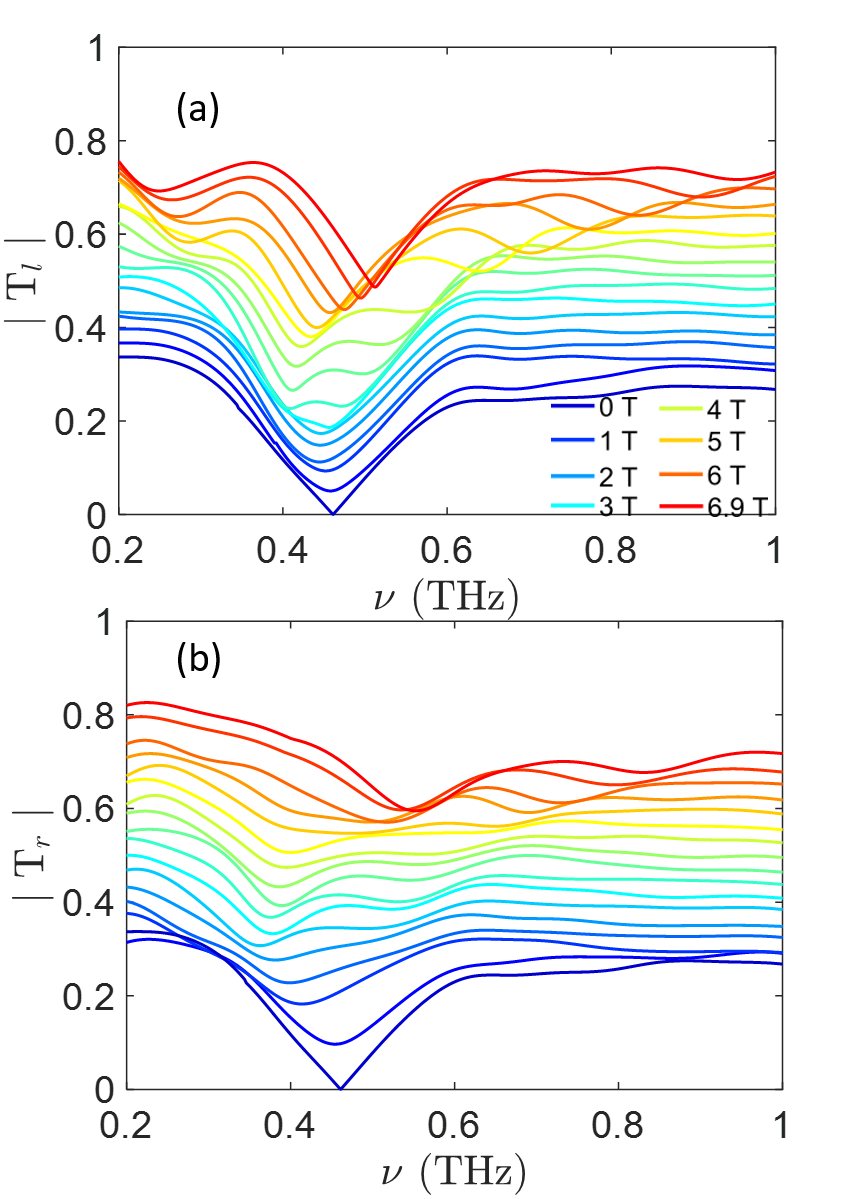}
	\caption{The transmission spectrum in \textbf{(a)} left and \textbf{(b)} right channel at different fields at 1.6 K}
	\label{fig:S2}
\end{figure}

The complex transmission of the sample in the linear basis is represented by a  $2\times2$ Jones matrix:

\begin{eqnarray}
\hat{T}=\begin{bmatrix}
T_{xx}       & T_{xy}  \\
T_{yx}       & T_{yy} \\
\end{bmatrix}
\end{eqnarray}

The transmission matrix of a sample with cubic symmetry in Faraday geometry is fully antisymmetric ~\cite{Armitage2014}, i.e. $T_{xx} = T_{yy}$ and $T_{xy} = - T_{yx}$. Therefore, the transmission matrix in circular basis can be diagonalized as

\begin{eqnarray}
\hat{T}_{cir}=\begin{bmatrix}
T_{xx}+iT_{xy}       & 0  \\
0       & T_{xx}-iT_{xy} \\

\end{bmatrix}=\begin{bmatrix}
T_{r}       & 0  \\
0       & T_{l} \\
\end{bmatrix},
\end{eqnarray}
where $T_{r}$ and $T_{l}$ denote the transmission matrix of right and left circularly polarized light respectively. With the polarization modulation technique as discussed in ~\cite{morris2012polarization}, $T_{xx}$ and $T_{xy}$ can be obtained simultaneously with the fast rotator method. This allows us to  determine the two complex transmission coefficients  $T_{xx}$ and $T_{xy}$ simultaneously and convert these to $T_{r}$ and $T_{l}$, which are the eigenpolarizations in the Faraday geometry.  The complex magnetic susceptibility can be calculated from the complex transmission as
$-\ln(T(\omega)) \propto \omega \chi(q=0,\omega)$.


\begin{figure}[h]
	\centering
	\includegraphics[width=0.98\columnwidth]{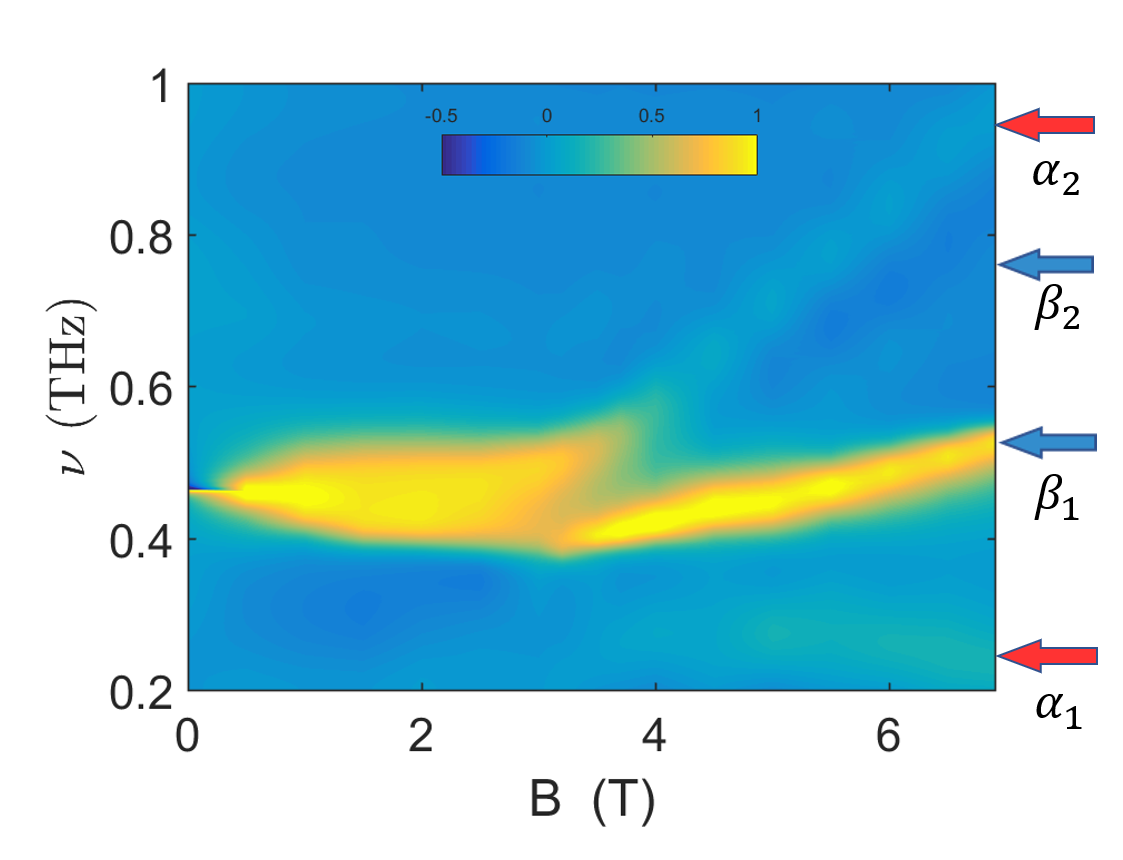}
	\caption{Image plot of ellipticity as function of magnetic field and frequency at 1.6 K}
	\label{fig:S13}
\end{figure}

\subsection*{3. Ellipticity}

As mentioned in the preceding section, we convert the transmission to the circular basis, i.e. the left and right channel as shown in Fig. 1(a) and (b) in main text. There are multiple excitations appearing in left and right channels. Ellipticity defined by $\eta = \frac{T_{l}-T_{r}}{T_{l}+T_{r}} $ puts excitations in left and right channels in one single plot.   It is nominally equivalent to the imaginary part of Faraday rotation tan($\phi_{F}$) $=\frac{T_{xy}}{T_{xx}}$. Ellipticity can amplify the difference between left and right circular basis and remove artifacts as it is a self-normalized quantity. As shown in Fig. S3, four branches are clearly seen, indicated by arrows. Ellipticity at zero field shows a singularity because the left and right polarized signals are identical at zero field as required by time-reversal symmetry.

\subsection*{4. CEF calculation and Steven operators in zero field }
The CEF Hamiltonian with D$_{3d}$ in the LS coupling scheme is given by
\begin{eqnarray}
H_{CEF}=\theta_{2}D_{2}^{0}O_{2}^{0}+\theta_{4}D_{4}^{0}O_{4}^{0}+\theta_{4}D_{4}^{3}O_{4}^{3} \nonumber\\+\theta_{6}D_{6}^{0}O_{6}^{0}+\theta_{6}D_{6}^{3}O_{6}^{3}+\theta_{6}D_{6}^{6}O_{6}^{6}
\end{eqnarray} where $O_{k}^{q}$ are the Stevens operators and $\theta_{k}$ are the reduced matrix elements: $\theta_{2}$ = -1/99, $\theta_{4}$ = 2/16335 and $\theta_{6}$ = -1/891891.  Crystal field parameters $B^{k}_{q}$ were obtained by fitting to zero field neutron crystal field energies ~\cite{PhysRevB.94.024430}.  $D^{q}_{k}$ and $B^{k}_{q}$ are related by $D^{q}_{k}=\lambda^{q}_{k}B^{k}_{q}$, where $\lambda^{0}_{2}=1/2$, $\lambda^{0}_{4}=1/8$, $\lambda^{3}_{4}=-\sqrt{35}/2$, $\lambda^{0}_{6}=1/16$, $\lambda^{3}_{6}=-\sqrt{105}/8$, $\lambda^{6}_{6}=\sqrt{231}/16$~\cite{PhysRevBcef}.  The lowest 13 CEF energy levels are the following:  0, 0, 1.9, 1.9, 10.4, 16.2, 44.1, 45.3, 45.7, 45.7, 62.9, 62.9, 69.8 meV. Note there are slight modifications of these numbers from the accepted values: the $\bf{k}$ averaged lowest excitation
1.6 meV was replaced by 1.9 meV at $\Gamma$ point as observed
in THz spectroscopy; the doublet at 44.1 meV is replaced by a doublet at 45.7 meV as confirmed by the authors in ~\cite{PhysRevB.94.024430}. The obtained $B^{k}_{q}$ are listed in Table 1 along with $B^{k}_{q}$ from Ref.~\cite{PhysRevB.94.024430} for comparison.
\begin{table}
 \begin{tabular}{|c| c c c c c c|} 
 \hline
&$ B^{2}_{0}$ & $ B^{4}_{0}$ & $ B^{4}_{3}$ & $ B^{6}_{0}$&$B^{6}_{3}$&$B^{6}_{6}$ \\ [0.5ex] 
 \hline\hline
 This work&52.2 & 302.0 & 129.3 & 65.6 &-31.5 & 149.8 \\

 \hline
From LS refined data~\cite{PhysRevB.94.024430} &55.9 & 310 & 114 & 64.6&-84.5&129 \\ [1ex] 
 \hline
\end{tabular}
\caption{Steven operators parameters (meV) $B^{k}_{q}$ obtained in this work and ~\cite{PhysRevB.94.024430} in LS coupling scheme. }
	\label{tab:1}
\end{table}

\subsection*{5. CEF calculation with Zeeman term in magnetic field }
The Zeeman energy is given by\begin{equation}
H_{Zeeman}=-\mu_{B}g_{J}\vec{B}\cdot\vec{J}
\end{equation} Since the magnetic field needs to be projected along local [111] direction, we need to do the calculation for Tb$_{\alpha}$ and Tb$_{\beta}$ separately because the Zeeman term is different for each due to relative orientation of the field with respect to local [111] axis.  We have $H_{Zeeman}^{\alpha}=-\mu_{B}g_{J}BJ_{z}$ and $H_{Zeeman}^{\beta}=-\mu_{B}g_{J}(BJ_{x}\mathrm{sin}\theta-BJ_{z}\mathrm{cos}\theta)$, where $\theta$ is the bond angle $\sim 109.5^{ \circ}$. By diagonalizing the matrix elements, we are able to obtain eigen-energies and eigen-functions at each field. As mentioned in main text, the comparison between experiments and calculation for Tb$_{\alpha}$ is reasonably good, while the calculated energies for Tb$_{\beta}$ are higher than the experimental values as shown in Fig. S4.

\begin{table*}
 \begin{tabular}{|c| c c c c c c |c|c c c c c c |} 
 \hline
D$_{3d}$&$ c^{2}_{0}$ & $ c^{4}_{0}$ & $ c^{4}_{3}$ & $ c^{6}_{0}$&$c^{6}_{3}$&$c^{6}_{6}$ & C$_{1}$ &$ c^{2}_{1}$& $ c^{2}_{2}$&$ c^{4}_{2}$&$ c^{4}_{4}$&$ c^{6}_{2}$&$ c^{6}_{4}$\\ [0.5ex] 
 \hline\hline
 Tb$_{\alpha}$&-0.46 & -1.5 & 0.9 & 0.1 &1.4 & -1.6&Tb$_{\alpha}$&0 &0 &0 &0 & 0& 0 \\ 
 \hline
Tb$_{\beta}$ &0.48 & -3.1 & -0.5 & -1&-0.3&1.9&Tb$_{\beta}$ &0.14&-0.21&-0.27&2.9&0.01  & 0.01  \\ [0.5ex] 
 \hline
\end{tabular}
\caption{The linear field coefficient $c_{q}^{k}$ for Tb$_{\alpha}$ with D$_{3d}$ and Tb$_{\beta}$ with C$_{1}$ symmetry in unit of meV.}
	\label{tab:1}
\end{table*}

\subsection*{6. Calculating intensity with Fermi's golden rule}
  According to Fermi's golden rule, the intensity should be proportional to the transition amplitude given by $I_{i\rightarrow j}\sim\frac{2\pi}{\hbar}\langle G_{1}\vert H^{'}\vert G_{2}\rangle ^{2}  $.  Here $H^{'}=\vec{B_{ac}}\cdot\vec{J}$ is  the time dependent perturbation from coupling between the ac magnetic field of THz pulses and magnetic dipole in Tb$_{2}$Ti$_{2}$O$_{7}$. In the Faraday geometry ($\bf{k}\!\parallel\!\bf{H}$)  along [111] direction, where $\bf{k}$ is the direction of light propagation, the transition matrix element $\langle i\vert H^{'}\vert j\rangle $ becomes $\langle i\vert B_{ac}J_{x}\vert j\rangle  $. $J_{x}=(J_{+}+J_{-})/2$, hence, the optical selection rule is  $\Delta m=+1$ for left and $\Delta m=-1$ for the right circular basis. Considering that we are at finite temperatures (1.6 K), the transition amplitude needs to be modified by the Boltzmann distribution.  Therefore, we have 
  \begin{equation}
  I_{i\rightarrow j}\sim |\langle i\vert J_{\pm}\vert j\rangle|^{2} \frac{exp[-E_{i}/k_{B}T]}{\sum\limits_{j}exp[-E_{j}/k_{B}T]}.
  \end{equation}\label{transition}
$J_{+}$ and $J_{-}$ are used for left and right circular transition, respectively.

\subsection*{7. Calculation with field dependent CEF }

\begin{figure}[h]
	\centering
	\includegraphics[width=0.8\columnwidth]{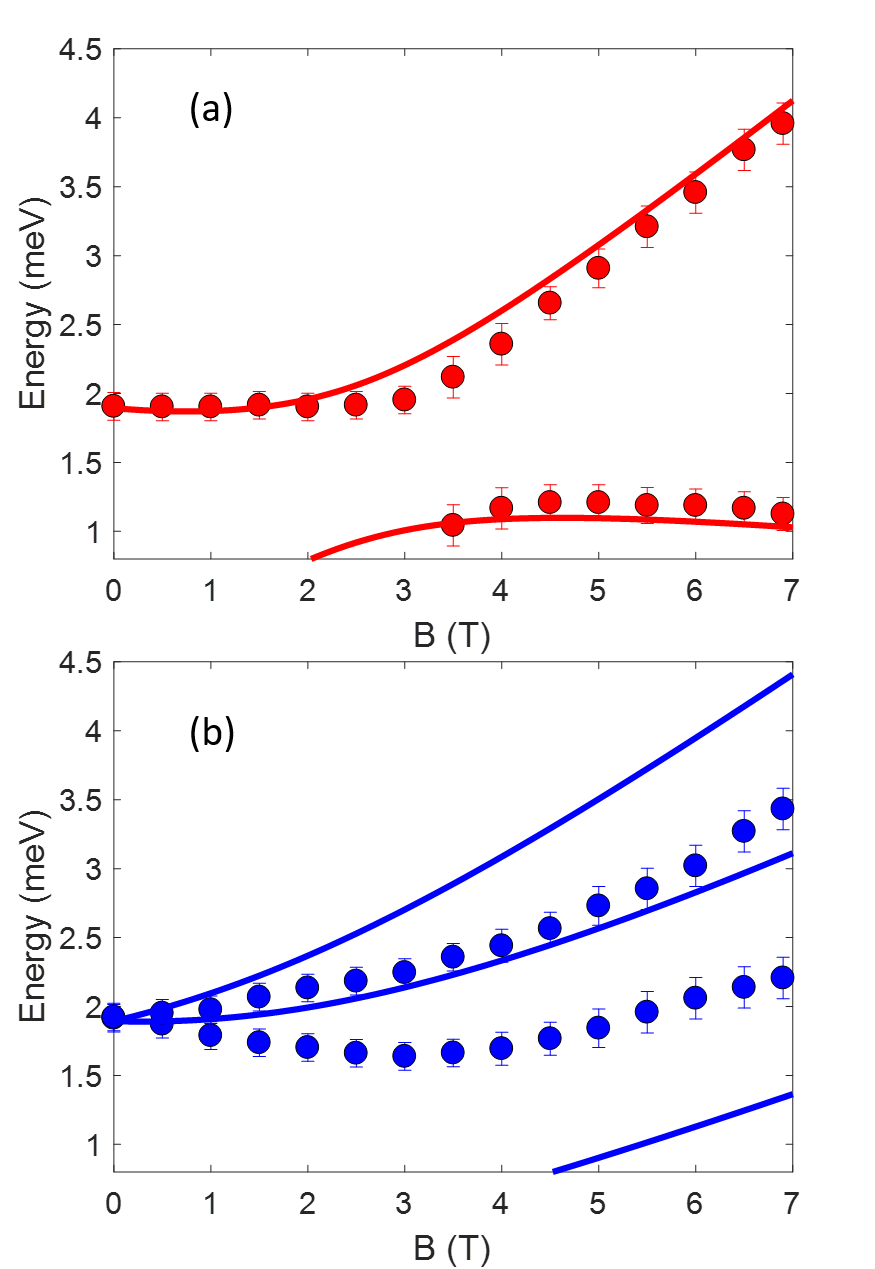}
	\caption{\textbf{(a)} Comparison between experiments and calculations with independent CEF of external field for Tb$_{\alpha}$ and \textbf{(b)}  Tb$_{\beta}$.  }
	\label{fig:S13}
\end{figure}

\begin{figure*}[t]
	\centering
	\includegraphics[width=2\columnwidth]{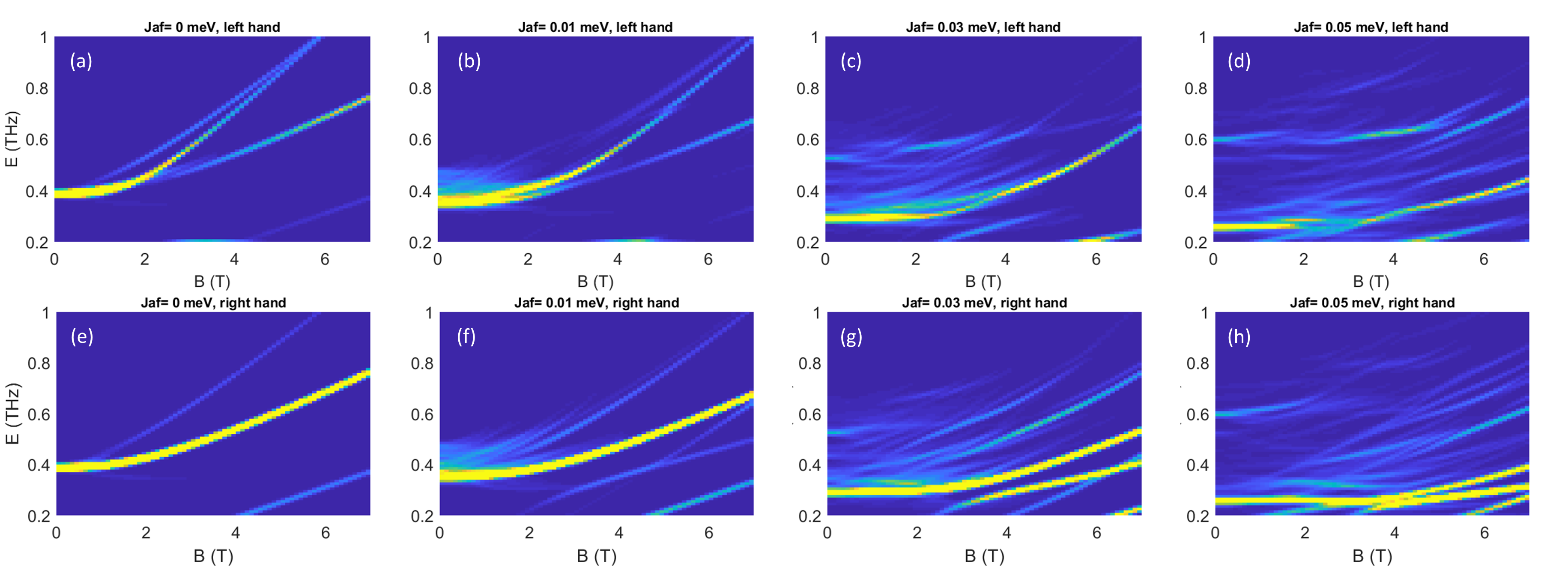}
	\caption{Simulated transmission amplitude for the full Hamiltonian $H_{\mathrm{int}}+H_{\mathrm{CEF}}+H_{\mathrm{Zeeman}}$. The corresponding $J_{\text{AF}}$ and handedness are labeled in the titles above (a-h). For this calculation, we keep the Hilbert space within only the four lowest CEF levels for each Tb ion and consider correlations within one tetrahedron, so the matrix of Hamiltonian of the dimension $4^4=256$. The wavefunctions of the kept CEF levels are taken from those of LS-coupling fitted results from Ref.~\cite{PhysRevB.94.024430}. The Lande g-factor is taken to be 1.5. The simulation gives a gap opening of the first CEF doublets ($\approx 1.8 \text{meV}\approx 0.4 \text{THz}$) and more complicated splittings of modes than the observed (see Fig. 1 of the main text) for $J_{\text{AF}}\gtrsim 0.02$ meV.}
	\label{fig:inter}
\end{figure*} 

Now we incorporate magnetic field induced CEF changes in our calculations. Applying field reduces the symmetry for Tb$_{\beta}$ from D$_{3d}$ to C$_{1}$, which allows additional Steven operators to appear in H$_{CEF}$.  It is sufficient to fit the data by including six of them: $O_{2}^{1}$, $O_{2}^{2}$, $O_{4}^{2}$, $O_{4}^{4}$, $O_{6}^{2}$, $O_{6}^{4}$.  On the other hand, Tb$_{\alpha}$ remains D$_{3d}$, so we keep the same Steven operators but add extra field dependence to the original $ B_{q}^{k}$. For simplicity, we approximate the dependence of the changes of CEF parameters to be linear on applied field, $ \Delta B_{q}^{k}=c_{q}^{k} B$, where $c_{q}^{k}$ is a coefficient with linear dependence.  Other dependencies could have been taken that would not have changed our ultimate result.  Therefore, the modified $B_{q}^{k}$ at finite field is sum of original $B_{q}^{k}$ at zero field and field induced change $\Delta B_{q}^{k}$. The relevant parameters are listed in Table 2. With increasing the external field further, the spin of Tb gradually aligns with magnetic field, so eventually there is a threshold of magnetic field $\sim$ 6 T, above which the CEF doesn't change any more. The total change of parameter $\Delta B_{q}^{k}$ is relatively small as compared to the original $ B_{q}^{k}$ in Table 1.

\begin{figure}[t]
	\centering
	\includegraphics[width=0.8\columnwidth]{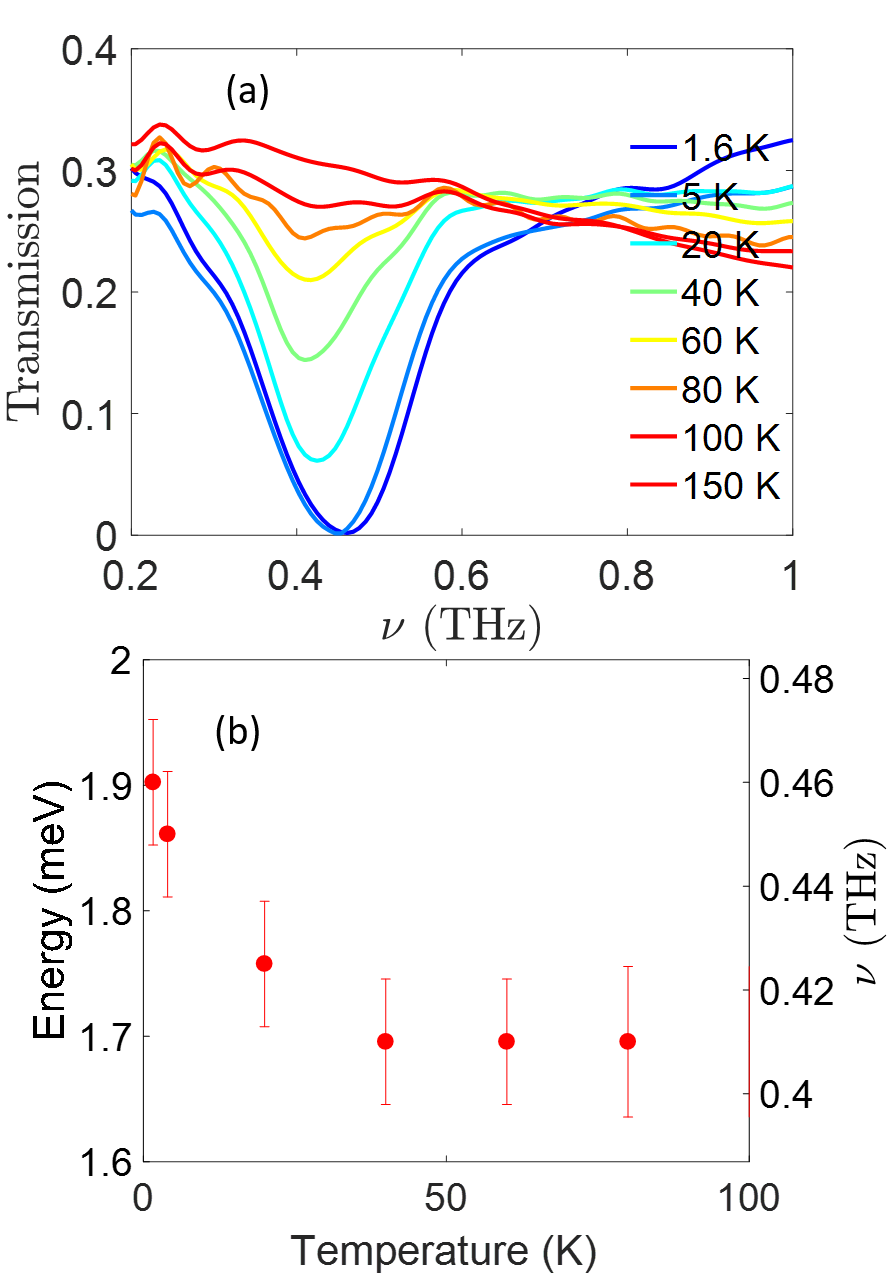}
	\caption{Transmission amplitude as a function of frequency
in zero field measured from 1.6 K to 150 K. (b) The central
frequency shifts slightly as a function of temperature. }
	\label{fig:S13}
\end{figure}

\subsection*{8. Exchange interactions}
In this section we show the influence of including antiferromagnetic interaction between the Tb spins on the simulated dispersion and transmission amplitude. Following Ref.~\cite{PhysRevLett.98.157204}, we consider isotropic antiferromagnetic interaction, namely
\begin{align}
H_{\mathrm{int}}=J_{\text{AF}}\sum_{\langle i,j\rangle}\mathbf{S}_i\cdot\mathbf{S}_j
\end{align}
besides $H_{\text{CEF}}$ and $H_{\text{Zeeman}}$. The summation of pairs $\langle i,j\rangle$ runs over Tb spins within a single tetrahedron, as the correlation length of Tb$_{2}$Ti$_{2}$O$_{7}$ at 1.6 K is estimated to be of the size of one tetrahedron~\cite{PhysRevB.62.6496}. With $J_{\text{AF}}$ gradually increased from 0 to 0.05 meV (Ref.~\cite{PhysRevLett.98.157204} estimates $J_{\text{AF}}\approx 0.167 \text{K}\approx 0.014 \text{ meV}$), the simulated transmission amplitude (by Eqn.\ref{transition} 5) evolves from Fig.~\ref{fig:inter} (a,e) to Fig.~\ref{fig:inter}(d,h). Compared to the observed transmission amplitude (see Fig. 1 of the main text), the calculated results show the opening of a gap between the first excited CEF doublets ($\approx$1.8 meV) at zero field and more complicated splittings of modes at finite field for $J_{\text{AF}}\gtrsim 0.02$ meV, while for smaller $J_{\text{AF}}$ the change from the result of $J_{\text{AF}}=0$ (no interaction) is not significant enough for a better agreement with the observation. Notice for this calculation we omit the dipolar interactions, which is estimated to be even smaller ($\approx$0.0315 K~\cite{PhysRevLett.98.157204}), and other interaction anisotropy. Despite the simplified model, we show the qualitative disagreement between the calculated results and the relatively simple dispersion observed. We conclude that the inclusion of a large exchange interactions are incompatible with the experimental result and unhelpful for a better fitting.

\subsection*{9. Magnetoelastic coupling}

Magnetoelastic coupling has been shown to exist in Tb$_{2}$Ti$_{2}$O$_{7}$ by many groups~\cite{fennell2014magnetoelastic,constable2017double,PhysRevB.99.224431}. We also find a shift of energy around 0.1-0.2 meV from the ground state to the first excited state potentially due to magnetoelastic coupling. The magnetoelastic coupling Hamiltonian has a form $H_{ME}=\sum\limits_{u,m} g_{u}\hat{U}_{u}\hat{O}_{n}^{m}$, where $\hat{U}_{u}=(\hat{a}_{u}+\hat{a}_{u}^{\dagger})$ is the phonon annihilation and creation operator with a phonon displacement $u$. $g_{u}$ is the magnetoelastic constant and $\hat{O}_{n}^{m}$ represents quadrupole operator. There are four quadrupole operators allowed to couple between E$_{u}$ transverse phonon and E$_{g}$ crystal field here, $\hat{O}_{2}^{2}=J_{x}^{2}-J_{y}^{2},  \hat{O}_{2}^{-2}=J_{x}J_{y}+J_{y}J_{x},  \hat{O}_{2}^{1}=J_{z}J_{x}+J_{x}J_{z},  \hat{O}_{2}^{-1}=J_{z}J_{y}+J_{y}J_{z}$. Therefore, we can calculate the matrix elements by using $J_{z}\vert j,m \rangle=m\hbar\vert j,m \rangle$ and $J_{\pm}\vert j,m \rangle=\sqrt{(j \mp m)(j \pm m+1)}\hbar\vert j,m \pm 1 \rangle$, where $J_{\pm}=J_{x}\pm iJ_{y}$. With a magentoelastic coupling  $g_{u}=0.006(2)$, our calculation shows the splitting of doublets is around 0.1-0.2 meV and may result in the shift of the energy as function of temperature as shown in Fig. S6. But such a magnetoelastic coupling only affect the spectra slightly and it is not a central ingredient to explain our data.

\subsection*{10. Specific heat measurements}

As shown in Fig. S7, the specific heat was measured at 1.6 K from 0 to 6 T.   Measurements were done in a Quantum Design PPMS. There is only a single peak at 0.7 T, which is related to the low energy CEF, and no signatures of a phase transition from 2-3 T. No signature of structural phase transition was observed.

\begin{figure}[h]
	\centering
	\includegraphics[width=0.98\columnwidth]{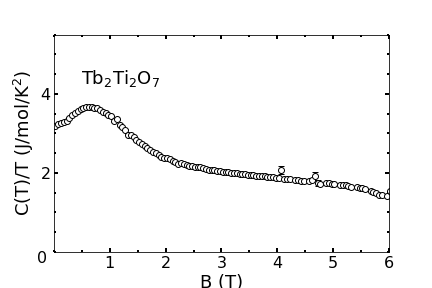}
	\caption{Specific heat was measured at 1.6 K from 0 to 6 T in a Quantum Design PPMS. There is a single peak at 0.7 T, which is related to low energy CEF, and no sharp peak around 3 T. No signature of structural phase transition was observed. }
	\label{fig:S13}
\end{figure}

\bibliography{referencesSI}